\documentclass[10pt,conference]{IEEEtran}
\let\labelindent\relax 
\pdfoutput=1
\usepackage{etex}

\usepackage[utf8]{inputenc}

\usepackage{mathptmx}
\usepackage[english]{babel}
\usepackage{multirow}
\usepackage{rotating}
\usepackage{color}
\usepackage{numprint}
\usepackage{xspace}
\usepackage{tabularx}
\usepackage{balance}
\usepackage{tikz}
\usepackage{pgfplots}
\usepgfplotslibrary{units}
\usepackage{caption}
\usepackage{framed}
\usepackage{mdframed}
\usepackage{latexsym}
\usepackage{amssymb}
\usepackage{amsmath}
\usepackage{comment}
\usepackage{subcaption}
\usepackage[inline]{enumitem}
\usepackage{xcolor}
\usepackage{algorithm}
\usepackage{algorithmic}

\usepackage{listings} 
\usepackage{textcomp}
\usepackage[bookmarks=false]{hyperref}
\usepackage{cite}
\usepackage{cprotect}

\usepackage{listings}
\lstdefinelanguage{diff}{
  language=java,
  basicstyle=\ttfamily\scriptsize,
  sensitive=true,
  morecomment=[f][\color{gray}][0]{diff},
  morecomment=[f][\color{gray}][0]{index},
  morecomment=[f][\color{blue}][0]{@@},
  morecomment=[f][\color{magenta}][0]{***},
  morecomment=[f][\color{violet}][0]{!},
  morecomment=[f][\color{red!60!black}][0]{-},
  morecomment=[f][\color{green!60!black}][0]{+},
  morecomment=[f][\color{magenta}][0]{---},
  morecomment=[f][\color{magenta}][0]{+++},
  morecomment=[f][\color{gray}][0]{Binary},
  morecomment=[f][\color{gray}][0]{Only},
  morecomment=[f][\color{gray}][0]{old},
  morecomment=[f][\color{gray}][0]{new},
  morecomment=[f][\color{gray}][0]{rename},
  morecomment=[f][\color{gray}][0]{similarity},
  morecomment=[f][\color{gray}][0]{deleted},
  morecomment=[f][\color{magenta}][0]{***************},
  morecomment=[f][\color{red!60!black}][0]<,
  morecomment=[f][\color{green!60!black}][0]>,
  morecomment=[f][\color{blue}][0]{0},
  morecomment=[f][\color{blue}][0]{1},
  morecomment=[f][\color{blue}][0]{2},
  morecomment=[f][\color{blue}][0]{3},
  morecomment=[f][\color{blue}][0]{4},
  morecomment=[f][\color{blue}][0]{5},
  morecomment=[f][\color{blue}][0]{6},
  morecomment=[f][\color{blue}][0]{7},
  morecomment=[f][\color{blue}][0]{8},
  morecomment=[f][\color{blue}][0]{9},
}[comments]

\usepackage[]{pdfcomment}

\newcommand{\TODO}[1]{\textcolor{red}{#1}}\newcommand\todo\TODO

\AtBeginDocument{

}

\newcommand\itzal{{Itzal}\xspace}
\newcommand\tool{Itzal4j\xspace}

\newcommand\productionApplication{Unmodified Application\xspace}
\newcommand\shadowService{Shadower\xspace}
\newcommand\patchService{Patch Synthesis Service\xspace}
\newcommand\regressionService{Regression Assessment Service\xspace}
\newcommand\requestOracle{Request Oracle Service\xspace}
\newcommand\reportingService{Patch Reporting Service\xspace}

\begin{document}
\title{Production-Driven Patch Generation}

\author{\IEEEauthorblockN{Thomas Durieux}
\IEEEauthorblockA{University of Lille \& Inria Lille, France}
\and
\IEEEauthorblockN{Youssef Hamadi}
\IEEEauthorblockA{Ecole Polytechnique, LIX, France}
\and
\IEEEauthorblockN{Martin Monperrus}
\IEEEauthorblockA{University of Lille \& Inria Lille, France}
}

\maketitle

\begin{abstract}
We present an original concept for patch generation: we propose to do it directly in production. Our idea is to generate patches on-the-fly based on automated analysis of the failure context. By doing this  in production, the repair process has complete access to the system state at the point of failure.
We propose to perform live regression testing of the generated patches directly on the production traffic, by feeding a sandboxed version of the application with a copy of the production traffic, the ``shadow traffic''. 
Our concept widens the applicability of program repair, because it removes the requirements of having a failing test case.
\end{abstract}

\IEEEpeerreviewmaketitle

\section{Introduction}

Program repair requires the presence of a failing test case to reproduce a failure that has happened in production. Writing such a failing test case is a really hard task, because the developer in charge of reproducing a failure has little access to the system state at the point of failure (she basically only has logs).
The difficulty of reproducing production failures has a direct impact on applicability of program repair: with no failing test, there is no patch generation. We aim at weakening the requirements of program repair by removing the mandatory presence of a failing test case.

Our intuition is to perform program repair directly in production, so that the repair process has a direct access to the system state at the point of failure.
This paper presents an architecture, called \itzal, it generates patches without requiring a failing test case.
The process of \itzal is as follows.
First, \itzal uses production assertions or runtime exceptions to detect failures.
Second, right after the failure is detected in production, a patch is searched in a sandboxed environment that mimics the production one. If a patch fixes the failure, it is a ``candidate patch''. 
Third, the patches are tested for regression, directly in production, based on traffic that is an exact copy of the production traffic -- we call it shadow traffic. 
\itzal has been realized in a prototype implementation for Java which focuses on generating source code patches for null dereferences.

This is a new line of research in automatic repair. Compared to classical test-suite based patch generation (e.g. \cite{le2012genprog}), \itzal does patch generation online, i.e. as soon as the failure happens, with no need for reproducing the failure. Yet, \itzal is not a classical runtime repair technique either (e.g. \cite{rinard2004enhancing}): while the patches are generated online in production, the system state is never altered. The \itzal patches are applied later, once the developer has validated them.

To sum up, our contributions are:
\begin{itemize}
\item \itzal, an architecture for patch generation in production that does not require a failing test case. 
\item The use of shadow production systems and shadow traffic in the context of automatic repair to generate patches in production.
\item The design and implementation of a Java implementation of this vision for null pointer exceptions. 
\end{itemize}

This paper is based on content from Arxiv's document \#1609.06848 \cite{durieux2016production} and is structured as follows.
\autoref{sec:contribution} presents \itzal.
\autoref{sec:rw} presents the related works and \autoref{sec:conclusion} concludes.

\section{\itzal{}}\label{sec:contribution}

\begin{figure}[t!]
\centering
\includegraphics[width=0.79\columnwidth]{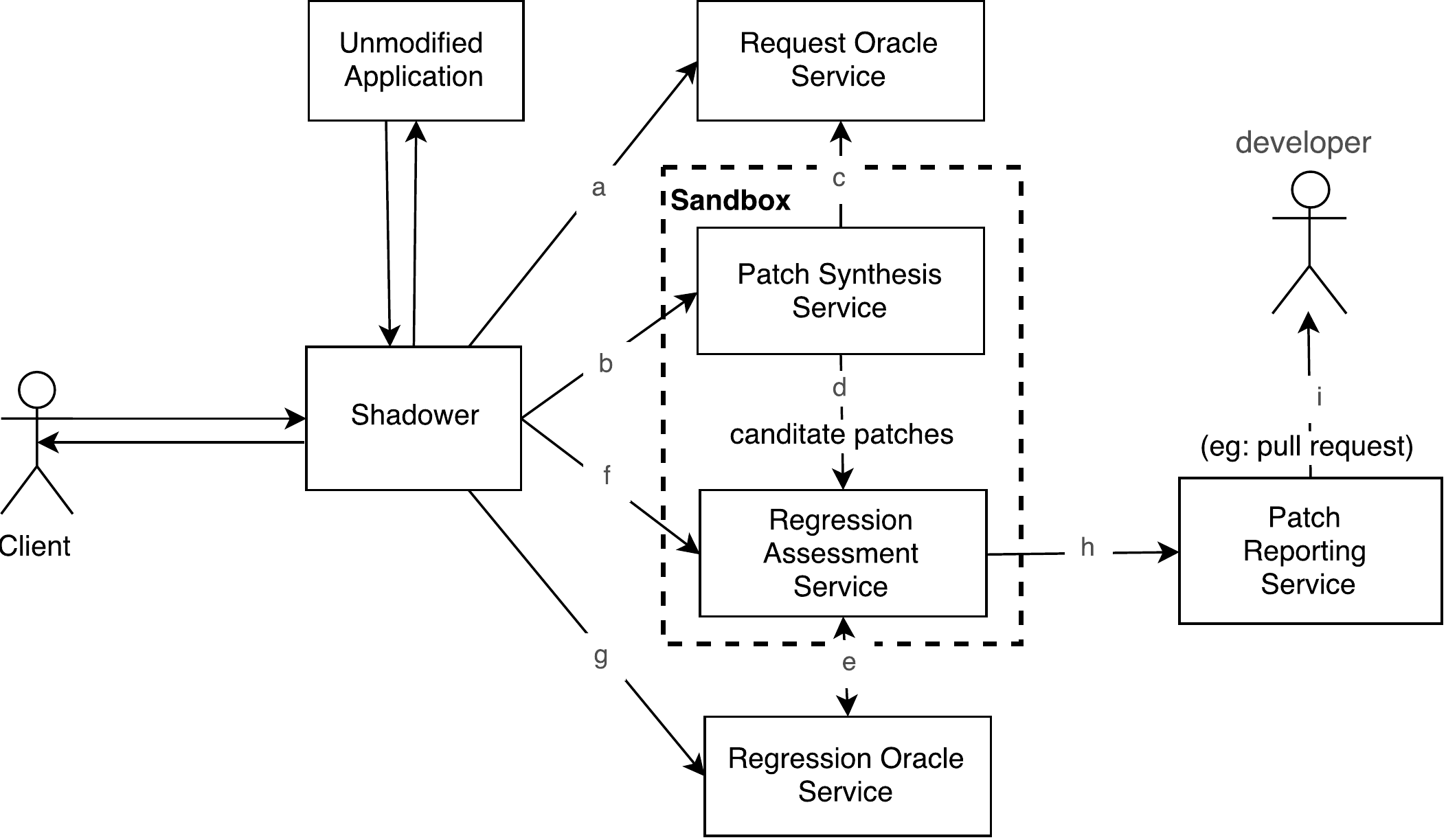} 
\caption{The Architecture of \itzal. The key idea is to duplicate production traffic via a ``shadower'', the duplicated traffic is used to search for patches and to validate candidate patches.}

\label{fig:architecture}
\end{figure}

We now present \itzal, a novel software repair technique for generating patches without requiring failing test cases, directly in the production environment. \itzal is a ``testless'' patch generation approach.

\subsection{Intuition}
The intuition behind \itzal is twofold. First, one can use production runtime contracts to drive the generation of source code patches. This includes classical pre- and post-conditions and implicit contracts such that an accessed variable must not be null. The latter is important because the violations of those implicit contracts come for free in any modern runtime, usually under the form of runtime exceptions. 

The second intuition is that one can use the diversity of the production inputs to perform in-the-field regression testing on the synthesized patches. This has the advantage that the regression exactly corresponds to what actually matters.

\subsection{Architecture}

The \itzal architecture is composed of seven components, as shown in \autoref{fig:architecture}.
\begin{enumerate}[leftmargin=*]
\item The \productionApplication (see \autoref{sec:application}) is the application onto which automatic patch generation is plugged. 
\item The \requestOracle (see \autoref{sec:request_oracle}) is a service that determines whether the application has successfully handle a request.

\item The \patchService (see \autoref{sec:patch_generation}) is the service that searches for patches that fix a given failure.

\item The \regressionService (see \autoref{sec:patch_regression}) performs regression testing on the generated patch. It applies the generated patches on the application and executes the request on it.

\item The Regression Oracle (see \autoref{sec:regression_oracle}) is the component that validates the generated patches by comparing the original output of \productionApplication to the output of the patched application for live production requests.

\item The \shadowService (see \autoref{sec:shadow}) is used to duplicate the requests of the \productionApplication. The duplicated requests are then sent in parallel to the \patchService and the \regressionService.

\item And the \reportingService (see \autoref{sec:patch_reporting}) is the component that selects the best patches and communicates them to the developers.
\end{enumerate}

\autoref{algo:main} shows the interactions between each component of the \itzal.
\itzal receives the request from the client (line \autoref{algo:shadow:request}).
Then it redirects the request to the \productionApplication (line \autoref{algo:shadow:send_application}).
Once the request has been handled by the \productionApplication, 
the response is sent back to the client, yet it is additionally sent to the \requestOracle   (arrow $a$ in \autoref{fig:architecture})  which verifies the viability of the output (line \autoref{algo:shadow:oracle}).
If the \requestOracle determines that there is a failure, the request is sent to the \patchService (arrow $b$ in \autoref{fig:architecture} and line \autoref{algo:shadow:send_patch}).
The patches generated by \patchService which pass the \requestOracle (i.e. fix the failure at hand) are sent to \regressionService (line \autoref{algo:shadow:push_patch}).
If the request has succeeded (no failure on the original application), the request is also sent to the \regressionService (line \autoref{algo:shadow:send_regression}) where all the previously generated patches are being validated on-the-fly against the new request.
When the \regressionService has identified valid patches with no regression, it sends them to the \reportingService.

To sum up, \itzal does patch generation online, i.e. as soon as the failure happens, directly in production. However, while the patches are generated online, they are applied later, once the developer has validated them. The side effects of patch search or regression testing on the production state are completely sandboxed, with no interference with the production environment.

\begin{algorithm}[t]
  \begin{algorithmic}[1]
    \REQUIRE{A: the \productionApplication}
    \REQUIRE{G: the \patchService}
    \REQUIRE{V: the \regressionService}
    \REQUIRE{O: \requestOracle}
    \WHILE{new request $r_{client}$ from Client}\label{algo:shadow:request}
        \STATE{$output$ = A($r_{client}$)} \label{algo:shadow:send_application}
        \STATE{send $output$ to Client}
        \IF{O($output$) is failure}\label{algo:shadow:oracle}
            \STATE{$patches$ = send $r_{client}$ to G} \label{algo:shadow:send_patch}
            \STATE{push $patches$ to V} \label{algo:shadow:push_patch}
        \ELSE
            \STATE{send $r_{client}$ to V for regression} \label{algo:shadow:send_regression}
        \ENDIF
        \IF{$\exists$ validated patches $\in V$}\label{algo:validation}
             \STATE{$p\leftarrow$ order the patches}
             \STATE{report $p$ to developers} \label{algo:report}
        \ENDIF
    \ENDWHILE
  \end{algorithmic}
  \caption{The main \itzal algorithm}
  \label{algo:main}
\end{algorithm}

\subsubsection{The \productionApplication} \label{sec:application}

\itzal augments a production application with automatic patch generation capabilities.
The requirement to deploy \itzal is that the application must use requests, ie. must have a message-driven architecture. A web application, or a web REST service are examples of message-driven applications.

The type of request may vary between applications, for example a request in a web application will be the request sent by a user's browser to a web-server, in a micro-service application, the request will typically be a REST message, in a mobile application, a request would be a touch event triggered when a user touches a mobile device's screen.

\subsubsection{\requestOracle} \label{sec:request_oracle}

The responsibility of the \requestOracle is to verify whether the application has succeeded to answer the request.
For instance, in a web-server, the \requestOracle can check the HTTP request return code (``assert response\_code != 500 (internal server error)''), of check the presence or not of an exception.
\itzal works with generic oracles such as checking the absence of exceptions (e.g. in a web request container or in a thread monitor), and it can also work with domain-specific oracles written by software engineers on top of domain concepts and data (e.g. the returned XML must comply with a specific schema).
When possible, the Request Oracle Service provide some information about the failure to help the \patchService to search for a patch. The information provided by default is the stack trace when the failure is based on an exception.

The \itzal does not require a perfect \requestOracle, i.e. the \requestOracle may miss some failures (false negatives).
In the case of false negatives, when the \requestOracle misses the failure detection, \itzal simply does not generate patches: this is unfortunate but it impacts neither the original application nor the patches generated for the other failures.
In the case of false positives, when the \requestOracle detects a failure when there is no failure in reality, the \patchService would generate patches, yet they would be lilkely benign if they pass regression testing done by the \regressionService.

\subsubsection{\patchService} \label{sec:patch_generation}

The \patchService is the service that synthesizes patches that fix a failing request.
\itzal can work with any patch synthesis approach compatible  with the \requestOracle.
The patches are applied to the failure point, for instance at the line where an exception has occurred. 
For a given failure point, the \patchService performs an exhaustive application of all possible patches.

For each tentative patch, \patchService calls the \requestOracle (arrow $c$ in \autoref{fig:architecture}) to verify that the request has been correctly handled by the patch template under consideration (the failure has been fixed). 
Because the \patchService generates the patches only based on one request (the failing one), the patches may break the behavior of the application for other requests, in other word, the may introduce a regression.
Thus, if the patch is successful on the failing request, the corresponding patch is transferred to the \regressionService (arrow $d$ in \autoref{fig:architecture}) that will further validates its correctness based on other requests.

The application and execution of candidate patches can change the state of the application in runtime.
Consequently, each execution is done in a sandboxed environment, this nullifies the potential side effects of the request or of the patch templates. 
The sandboxed environment contains a shadow state of the application, which is regularly synchronized with the production one.
Since the space of patches is sometimes large, \itzal uses a time budget.
It explores the patch alternatives sequentially until they are all explored or until the time budget is consumed.

\itzal can work with any patch model, whether domain-specific (such as out-of-bounds exception) or generic (à la Genprog \cite{le2012genprog}).
Similarly, \itzal can be applied to binary code if the patch synthesis technique supports it. 
Our current prototype generates patches for null dereferences.
If the patch model generates too much patches, i.e. the search space is too large, this would be a problem because it would represent a huge computation effort on the \patchService and much more importantly on the \regressionService.

\subsubsection{\regressionService} \label{sec:patch_regression}

The patches generated by the \patchService can introduce regressions because their generation only involves one request (the failing one).
The \regressionService has the responsibly to check the behavior of the application when the generated patches are injected against other requests.
It detects these regressions by comparing the output of the \productionApplication against the output of the patch-augmented application.
If the output is different, 
it means that the patch has introduced a regression, and  the patch is consequently marked as invalid.
This comparison is done on-the-fly, directly on production traffic. 
Doing regression testing ``live'' has the advantage that there is no need to record the potentially enormous amount of production data.

\subsubsection{Regression Oracle}\label{sec:regression_oracle}
The Regression Oracle compares the output of the \productionApplication (arrow $g$ in \autoref{fig:architecture}) and the output of a patched version in the \regressionService for the same request.
If the outputs are different, the Regression Oracle marks the current patch as invalid.
For example, a regression oracle for a web server compares the HTML text of both versions.
The comparison is not necessarily a byte-to-byte one, it can include heuristics to discard transient information such as time, cookie identifiers, etc.

\subsubsection{\shadowService} \label{sec:shadow}

The role of the \shadowService is to create shadow traffic from actual end-user traffic coming into the application.
The ``shadow traffic'' is made of production requests that are duplicated one or several times and sent to sandboxed shadow applications.
In our case, the shadow applications are the \patchService and \regressionService.

In \itzal, the \shadowService receives the requests from the clients duplicates them and sends one duplicate to each service of the architecture  (arrows $a$, $b$, and $f$ in \autoref{fig:architecture}). The response is also shadowed for the regression oracle service (arrow $g$ in \autoref{fig:architecture}).

In the context of web applications, the concept of running multiple instances of an application is well known and heavily used: this is done for load balancing and rolling deployment. 
The difference between a load balancer and a \shadowService is twofold:
first, a load balancer does not duplicate the traffic;
second, a load balancer does not send requests to sandboxed "sinks" as \itzal does.

Since \itzal is a production technique, it must have a reasonable impact on the performance of the application.
In order to minimize the impact on the \productionApplication, \itzal computes the \regressionService and the \patchService asynchronously.
Indeed, the goal of \itzal is to perform patch generation, not automatic error recovery system.
Hence, the \shadowService directly sends the output as soon as the  \productionApplication has handled a request (even if there is a failure).
\itzal does not have to wait for the end of the patch search or the end regression testing for sending the response back to the client.
The \shadowService is thus the only component that impacts the performance of the \productionApplication. 
The \shadowService requires to 1) copy and reroute requests on the fly and 2) maintain an appropriate shadow state of the system under consideration.
In a typical HTTP-based setup, the cost of the former is similar to that of classical web proxies and load balancers, which are extensively used in production systems.
The latter point is more an open question: very few works study production state shadowing, neither in academia nor in industry. It may be a costly operation if databases are naively copied for instance. However, we envision piggy-backing on the latest advances in efficient online backups and copies of file systems.

\subsubsection{\reportingService} \label{sec:patch_reporting}

The \reportingService is the service that communicates the results of \itzal to the developers (arrow $i$ in \autoref{fig:architecture}).

It happens that multiple patches (corresponding to multiple patch templates) successfully pass the regression test over production traffic.
Consequently the \reportingService has to sort the patches in order to first propose the most useful ones to the developers.
To sort, the \reportingService uses the number of execution of the patched line in the \regressionService (the number of requests that execute the patch). The idea is that the more a patch has been executed by the \regressionService, the less likely it is to introduce a regression.

We now discuss the reporting medium to the human developer. There are several types of communication that can be used in the \reportingService.
In the current prototype, we have a dashboard where the developers follow in real time the failures, the generated patches and the progression of the patch validation.

\subsection{Prototype Implementation for Java}

We have implemented a prototype of \itzal for Java in a tool named \tool, dedicated to reactive applications based on HTTP.
\tool generates patches for null dereference failures.
In \tool,  the \requestOracle is based on exceptions.
Any uncaught exception happening during the processing of a request is considered as a failure.
The \patchService is dedicated to null pointers and uses the NPEFix  technique \cite{cornu:hal-01251960} for searching the space of possible patches for null dereferences.
Sandboxing of patch search is achieved using Docker, a major software containerization platform which provides powerful sandboxing (both disk and IO based).
In our implementation, the \patchService sends all candidate patches to the \regressionService using an HTTP-based protocol.
For the Regression Oracle, we compare the body of the HTTP response of the \productionApplication against the output produced by the patched application (e.g. the HTML body text). 
If the outputs match, the patch is considered validated for the current request otherwise the patch is permanently marked as invalid.
The \shadowService is implemented on top of a HTTP proxy implementation in Java called ``Jetty Proxy''.
The \reportingService of \tool is a web  dashboard, where the developers can access in real time the current patches of \itzal: the ones that have fixed at least one failure and the patch that are under regression testing.
For each patch, they can visualize the number of failures of the system detected by the \requestOracle, see the actual patch code, and the patch validation metrics such as the number of executions done by the \regressionService.

\section{Related Work}
\label{sec:rw}

Our work is much inspired by the classical work on runtime repair.
Rinard et al. \cite{rinard2004enhancing} present a technique called ``failure oblivious computing'' to avoid illegal memory accesses by adding additional code around each memory operation during the compilation process.
Assure \cite{sidiroglou2009assure} is a self-healing system  based on error-virtualization.
Long et al. \cite{long2014automatic} proposes the concept of ``recovery shepherding'' in a system called RCV.
Those techniques do not produce patches and do not perform regression testing in production.

Gu et al. \cite{gu2016automatic} presents Ares a runtime error recovery for Java exceptions using JavaPathFinder (JPF).
The two majors differences with \tool are: first \tool is safer, it does not modify the production state  of the application as Ares does, and secondly, while Ares performs runtime repair, \tool produces source code patches that are then communicated to the developers. 

Perkins et al. \cite{perkins2009automatically} propose ClearView, a system for automatically repairing errors in production.
\itzal and ClearView both perform repair in production, yet they are very different:
1) ClearView does not produce source code patches while Itzal does;
2) ClearView modifies the production state, while Itzal only modifies the sandboxed shadow requests and state (this means that ClearView can mess up the application while Itzal never does so);
3) ClearView works with learned invariant-based oracles, while Itzal uses human designed request oracles.

The concept of shadow traffic is related to the execution of multiple versions of the same software in parallel, called in the literature ``multi-version execution'' \cite{hosek2013safe},  or ``parallel execution'' \cite{Trachsel2010ParallelExecution}.
However, none of the related work uses shadow traffic to generate patches.

\section{Conclusion}
\label{sec:conclusion}

In this paper, we have presented \itzal, an approach for generating patches in production. The failure detection that triggers the patch search is achieved with runtime assertions, and the regression assessment is done on live production traffic. 
In \itzal, the patch search is done in a fully sandboxed environment, with no interference with the production data.
Our future work now consists of devising an approach to fully automatically store a shadow of the production state, and to efficiently synchronize the shadow state with the actual production state.

\balance

\bibliographystyle{IEEEtran}
\bibliography{references}

\begin{thebibliography}{1}
\providecommand{\url}[1]{#1}
\csname url@samestyle\endcsname
\providecommand{\newblock}{\relax}
\providecommand{\bibinfo}[2]{#2}
\providecommand{\BIBentrySTDinterwordspacing}{\spaceskip=0pt\relax}
\providecommand{\BIBentryALTinterwordstretchfactor}{4}
\providecommand{\BIBentryALTinterwordspacing}{\spaceskip=\fontdimen2\font plus
\BIBentryALTinterwordstretchfactor\fontdimen3\font minus
  \fontdimen4\font\relax}
\providecommand{\BIBforeignlanguage}[2]{{%
\expandafter\ifx\csname l@#1\endcsname\relax
\typeout{** WARNING: IEEEtran.bst: No hyphenation pattern has been}%
\typeout{** loaded for the language `#1'. Using the pattern for}%
\typeout{** the default language instead.}%
\else
\language=\csname l@#1\endcsname
\fi
#2}}
\providecommand{\BIBdecl}{\relax}
\BIBdecl

\bibitem{le2012genprog}
C.~Le~Goues, T.~Nguyen, S.~Forrest, and W.~Weimer, ``Genprog: A generic method
  for automatic software repair,'' \emph{IEEE Transactions on Software
  Engineering}, vol.~38, no.~1, pp. 54--72, 2012.

\bibitem{rinard2004enhancing}
M.~C. Rinard, C.~Cadar, D.~Dumitran, D.~M. Roy, T.~Leu, and W.~S. Beebee,
  ``Enhancing server availability and security through failure-oblivious
  computing.'' in \emph{OSDI}, vol.~4, 2004, pp. 21--21.

\bibitem{durieux2016production}
T.~Durieux, Y.~Hamadi, and M.~Monperrus, ``Production-driven patch generation
  and validation,'' \emph{arXiv preprint arXiv:1609.06848}, 2016.

\bibitem{cornu:hal-01251960}
T.~Durieux, B.~Cornu, L.~Seinturier, and M.~Monperrus, ``Dynamic patch
  generation for null pointer exceptions using metaprogramming,'' in \emph{IEEE
  International Conference on Software Analysis, Evolution and Reengineering},
  2017.

\bibitem{sidiroglou2009assure}
S.~Sidiroglou, O.~Laadan, C.~Perez, N.~Viennot, J.~Nieh, and A.~D. Keromytis,
  ``Assure: automatic software self-healing using rescue points,'' \emph{ACM
  SIGARCH Computer Architecture News}, vol.~37, no.~1, pp. 37--48, 2009.

\bibitem{gu2016automatic}
T.~Gu, C.~Sun, X.~Ma, J.~L{\"u}, and Z.~Su, ``Automatic runtime recovery via
  error handler synthesis,'' in \emph{Proceedings of the 31st IEEE/ACM
  International Conference on Automated Software Engineering}.\hskip 1em plus
  0.5em minus 0.4em\relax ACM, 2016, pp. 684--695.

\bibitem{perkins2009automatically}
J.~H. Perkins, S.~Kim, S.~Larsen, S.~Amarasinghe, J.~Bachrach, M.~Carbin,
  C.~Pacheco, F.~Sherwood, S.~Sidiroglou, G.~Sullivan \emph{et~al.},
  ``Automatically patching errors in deployed software,'' in \emph{Proceedings
  of the ACM SIGOPS 22nd symposium on Operating systems principles}.\hskip 1em
  plus 0.5em minus 0.4em\relax ACM, 2009, pp. 87--102.

\bibitem{hosek2013safe}
P.~Hosek and C.~Cadar, ``Safe software updates via multi-version execution,''
  in \emph{Proceedings of the 2013 International Conference on Software
  Engineering}.\hskip 1em plus 0.5em minus 0.4em\relax IEEE Press, 2013, pp.
  612--621.

\bibitem{Trachsel2010ParallelExecution}
O.~Trachsel and T.~R. Gross, ``Variant-based competitive parallel execution of
  sequential programs,'' in \emph{Proceedings of the 7th ACM International
  Conference on Computing Frontiers}, 2010.

\end{thebibliography}

\balance
\end{document}